# Electrical transport, magnetic, and structural properties of the vortex lattice of $V_3Si$ in the vicinity of the peak effect

**A. A. Gapud[*,a], D. K. Christen[*], J. R. Thompson[*,+], and M. Yethiraj[*]**

[*]*Oak Ridge National Laboratory, Oak Ridge, TN 37831-6061 USA*

[+]*Department of Physics, University of Tennessee, Knoxville, TN 37996-1200 USA*

## Abstract

The peak effect in critical current density $J_c$ is investigated by studying the flux dynamics in $V_3Si$ using bulk magnetometry, small-angle neutron scattering, and transport measurements on clean single-crystal samples from the same ingot. For a field-cooled history, well-defined structure in the vortex lattice was found for fields and temperatures *(H,T)* below the peak-effect line $H_P(T)$; above this line, the structure disappeared. History-dependent, metastable disorder is found only for *(H,T)* below $H_P(T)$ but the vortex system is reproducibly re-ordered either by field-cooling or a low-frequency, pulsed "shaking" transport current. The latter is shown to attain Bardeen-Stephen flux flow. In addition, flux flow is observed at $H_P(T)$ at high current levels. The results support the traditional picture of $H_P(T)$ as an order-disorder transition due to the competition between elasticity and pinning.

[a]corresponding author, email address: gapudaa@ornl.gov

## I. INTRODUCTION

The peak effect – in which the critical current density $J_c$ exhibits a local maximum for weak-pinning superconducting systems just below the upper critical field line $H_{c2}(T)$ – is currently a topic of renewed interest. Recent studies have revealed that the peak effect occurs in all superconductor species, including the newly discovered $MgB_2$ [1], leading to open discussion of a unifying magnetic phase diagram $H(T)$ for both low- and high-temperature superconductors. These studies focus specifically on the dynamics of vortices about the peak-effect line $H_P(T)$, where in many cases the situation is simplified by minimizing thermal effects and anisotropy effects through the use of clean single-crystal samples of isotropic, low-temperature superconducting species, specifically Nb, $V_3Si$, and $(K,Ba)BiO_3$. Therein, the most popular methods used have been magnetization measurements [2], transport measurements [3], and small-angle neutron scattering (SANS)[4 – 10] on various samples. In this present work on $V_3Si$ all three of these complementary measurement methods were utilized, and were performed on clean, weak-pinning samples from the same single crystal ingot.

This study provides additional confirmation of the traditional picture of the peak effect as an order-disorder transition, as was originally formulated by Pippard and by Larkin and Ovchinnikov [11] long before the advent of high-temperature superconductors. The model describes a competition between the elastic energy of the vortex lattice and the pinning energy at fields and temperatures close to the transition to normal state. In this region, the softening of the lattice elasticity leads to more effective pinning. Since the pinning disorders the lattice, the peak effect has been seen as an order-disorder transition line. This transition has been observed recently in other cubic species, $(K,Ba)BiO_3$ [6,7] and Nb [9], wherein a quasi-ordered Bragg glass is thought to cross over to a purely disordered state. A more recent confirmation utilized specialized scanning tunneling microscopy [12] which revealed a clear transition from a collective, orderly vortex-lattice to uncorrelated, disorderly motion upon crossing $T_p(H)$. These features also confirm the results of a notable numerical simulation [13] modeling the encroachment of disorder as pinning overwhelms lattice elasticity as field and temperature are increased.

A recent concern regarding the pinning-versus-elasticity picture is the apparent complication of history effects in the preparation of the vortex lattice which have been ascribed to a sub-peak *metastable phase* [2, 9]. For the regime examined in this study, it will be shown that such metastability can be reproducibly minimized.

## II. EXPERIMENT

The samples were all taken from a clean (RRR = 39) single-crystal of $V_3Si$ with high Ginzberg-Landau parameter $\kappa$ of 25. The cylindrical, mother ingot was the same crystal used in Ref. 14, with an estimated mean free path $l$ = 320 Å which is much larger than the coherence length $\xi_o$ = 38 Å, a critical temperature $T_c$ of 16.3 K, and thermodynamic critical field $H_c$ of about 0.5 T. The Ginzburg number *Gi*, which quantifies the significance of thermal-fluctuation effects, is estimated to be less than $10^{-6}$, which would *a priori* eliminate any measurable possibility of a "melting" transition distinct from the superconducting transition [15].

All measurements were done with the field applied parallel to the crystallographic <110> direction. The neutron scattering was conducted using horizontal-field geometry at various constant fields as a function of temperature and thermal history. Magnetization was measured on a disk section using a SQUID magnetometer, with the field along the central axis and as a function of applied magnetic field and temperature. A 1-mm square-cross-section rod sample was used for transport *V(I)* measurements using a standard, 4-contact strip configuration of negligible contact resistance, with the field perpendicular to the length of the rod. The contacts for the current consist of indium pads which were ultrasonically soldered onto the rod, yielding a contact resistance below $10^{-6}$ Ω. This enables a maximum system current level of more than 50 A, although in this study there was almost no need to exceed 10 A. The applied "shaking" current was a low-frequency rectangular wave where each *cycle* is followed by a null signal of 2.5 sec. For currents ≤ 100 mA, the cycle is the 2-second period of a rectangular wave. For currents > 100 mA the current is applied over a much shorter interval of 30 ms, with each pulse separated by a null signal of 0.5 sec. For all currents, voltage and current readings are sampled over a smaller interval, 17 ms, within each pulse, and the final readings are averaged over one to three cycles.

## III. RESULTS AND DISCUSSION:

A first signature of the peak effect comes from the magnetic hysteresis measurements, plotted in Fig. 1. The weak pinning in the system is evident from the rapid collapse of low-field irreversibility into an almost-reversible regime at higher fields, developing a subtle "fishtail" just below the normal-state transition field $H_{c2}(T)$. Using the Bean model, the corresponding $J_c$ peak is shown in the lower plot, where peaks for two other temperatures are also shown. These features are qualitatively similar to the results of another study on high-RRR samples [2], making the samples in the present study an ideal system for observing the peak effect. These features are also seen in the temperature domain for several fields, although the hysteresis is below instrument resolution at lower fields; Fig. 2 shows the case for $H$ = 6.0 T, where the $J_c$ peak appears at temperature $T$ = 13.0 K.

The observed collapse in magnetic irreversibility has been interpreted as a complete "melting" into a vortex liquid deep within the mixed state. However, in this same $H$-$T$ regime our SANS studies of of the vortex matter show the existence of long-range order, or a significant vortex-lattice phase. Fig. 3 shows the intensity of a primary Bragg spot for the sample field-cooled (FC) at $H$ = 6.0 T and then subsequently warmed (FCW). The extent of order shows saturation at lowest temperatures and decreases monotonically towards the normal-state transition $T(H_{c2})$ ~ 13.7 K determined from the magnetization measurements. However, all ordered structure disappears at a slightly lower temperature, 13.0 K, which according to Fig. 2 is the temperature $T_p$ (6 T) where the peak effect occurs.

Consistent with this is the clear radial and azimuthal broadening of the same Bragg spot when approaching $T_p$, as shown in the lower plots. The data also shows history effects: the SANS intensity is noticeably higher for the vortex lattice after it has been cooled to base temperature and then warmed, implying a type of lattice-defect "healing" at lowest temperatures. The ordering effect of cooling is also seen in the peak breadth, which is narrower upon warming.

These observations are similar to those in other cubic superconductors, $(K,Ba)BiO_3$ [7] and Nb [9]. Consistent with Ref. 9 is our observation of Bragg-spot broadening and history effects. While the result and interpretation for Nb is controversial [10], in (K,Ba)$BiO_3$ the regime below $T_p(H)$ is interpreted as a quasi-lattice or *Bragg glass* [7,8] where disorder was shown to be significantly reduced by adding a small oscillatory component to the field while cooling. This is also reminiscent of the well-known order-disorder coexistence seen for the corresponding regime in anisotropic 2H-$NbSe_2$ [16,17,18].

The observation of both significant Bragg neutron scattering in a field-cooled vortex lattice and the near-reversibility of magnetization raises the possibility of observing Bardeen-Stephen flux flow (BSFF), where the elasticity of an ordered moving lattice could dominate weak pinning. The simple ratio between flux-flow resistance $\rho_f$ and normal-state resistivity $\rho_N$ is given by [19]

$$\frac{\rho_f}{\rho_N} = C\frac{H}{H_{c2}} = CH\left[\left(\frac{dH_{c2}}{dT}(T_c)\right)(T - T_c)\right]^{-1} \qquad (1)$$

where $C$ is a dimensionless factor of order unity in this regime and we use $dH_{c2}/dT\ (T_c)$ = - 2.3, consistent with our final data (Fig. 10).

A direct way to examine the possibility of BSFF is through the direct application of transport current and observing the voltage-current $V(I)$ response in light of equation (1). Doing this at various points $(H,T)$ about $H_p(T)$ (determined from magnetization measurement) is therefore a way of probing for a transition between order and disorder as a result of competition between elasticity and pinning, and whether or not this indeed occurs at $H_p(T)$ – i.e., the Pippard-Larkin-Ovchinnikov picture. This is almost analogous to a recent study on $NbSe_2$ [18] wherein the metastable phase is seen as "blurring" the location of the order-disorder transition; when the metastability was significantly minimized, this transition was shown to occur exactly on the peak-effect line $H_p(T)$. In the present study, the fact that BSFF could be achieved at all, without contact heating or driving the sample into the normal state first, is further testament to the quality of the samples in this study.

The first exercise is to compare $V(I)$ curves generated in two different ways: (a) Increasing $I$: preparing a virgin lattice and then applying pulsed current where the pulse height $I$ starts at lowest levels – lower than critical current $I_c$ – and is then progressively increased; (b) Decreasing $I$ : starting $I > I_c$ at highest values (~ 10 A), then progressively lowering to lowest levels. The presumption is that scheme (a) would probe the virgin lattice along with any metastable disorder, while scheme (b) would *ab initio* drive out any such disorder before the curve is traced. These are shown in Fig. 4 for various temperatures at a constant field 4.3 T, for which $T_p$ was identified as 14.0 K. We observed the following: (1) For all $T$, the $V(I)$ generated by scheme (b) is thereafter independent of the history by which $I$ is changed; i.e., it becomes reversible or "stable" with respect to $I$, as therefore indicated by the double arrow. (2) For $T < T_p$ (main panel and top inset) the $V(I)$ generated by scheme (a) has a higher $I_c$ than that generated by scheme (b) and eventually joins with the latter at higher $I$, but (3) for $T \geq T_p$, the two $V(I)$ curves coincide, as shown in the two lower insets. This was found to be typical over similar other sets of $(H_p,T_p)$. It clearly shows that the metastability occurs only for $(H,T)$ below the peak effect line $H_p(T)$ and could yield an anomalously larger $I_c$ and, in turn, an anomalous $J_c$ peak as also seen in recent transport experiments with $NbSe_2$ [18,20 – 23]. Scheme (b) therefore provides a reliable procedure for determining a stable $I_c(H,T)$.

We note further that for $T < T_p$ the $V(I)$ generated by scheme (a) is dependent on the history of $I$ ; this is shown in Fig. 5, where the measurement is stopped at progressively higher $I = I_s$ before starting over at the same minimum current. Each time the measurement is restarted, a different curve is generated. The first curve has the highest $I_c$, but as $I_s$ is increased, the curve generated is progressively closer to the stable, low-$I_c$ curve, until eventually becoming reversible with $I$. This is further illustrated in the inset of Fig. 5 by the sharp decrease of $I_c$ to a saturation value as $I_s$ is increased. A subtlety is also noted in this plot: While the stable $V(I)$ is not yet reached, there is a tendency to return to the original virgin-lattice curve: the lowest three curves seem to merge at higher currents into the maximum-$I_c$ curve of Fig. 5. $I_s$ always ends up on this curve, except after it ends up on the stable curve, after which the stable curve is retraced for any $I_s$ . This behavior is strongly suggestive of a "settling" back into the virgin condition, as a metastable ground state. Such would be the case if weakly driven vortices were to accommodate themselves into random pinning sites.

The approach towards the stable curve can also be achieved at a constant $I$. This is shown for two different current levels $I$ in Figs. 6 (a) and (b). For each current level, we start with either a stable $V(I)$ or one for a virgin lattice, then ramp $I$ up to a constant level, where we then apply the pulsing current over the time indicated, and then continue ramping beyond. The time development shown in the inset underscores the distinction between stable and metastable *V(I)* in this exercise: in the former case, the voltage is constant over time – which also confirms that contact heating is negligible – while for the latter there is a gradual approach to the stable voltage. Fig. 6(b) is notable in that the current level $I$ is such that after about one hour, the *V(I)* has reached its stable level and upon resumption of $I$ ramping, the stable curve is traced out. Meanwhile, in Fig. 6(a), the stable curve is not reached during the same time and shows the "settling" behavior discussed in Fig. 5: upon resumption of $I$ ramping the curve flattens out to re-merge to the original *V(I)* indicated by the dashed curve.

The most important feature of Figures 4 to 6 is that for $T < T_p$, each of the stable *V(I)* curves contains a broad portion which is ohmic (slope=1 in a log-log plot). This is highly suggestive of dissipative flux flow preceded by disordered creep-like flow at smaller driving amplitudes $I$. The ohmic behavior is indicated by a dashed straight line in each plot. In order to confirm BSFF, stable *V(I)* curves both above and below the peak effect $(H_p, T_p)$ were obtained at various *(H,T)*. A typical set is shown in Fig. 7 for a constant temperature of 14.0 K and fields from 3.5 to 5.5 T. The fields are in increments of 0.1 T, except for the two adjacent to $H_p$ with steps of 0.05 T. (The curves in Fig. 7 were measured with voltage taps which were 25% closer together than those used for the data in Fig. 4, thus yielding slightly lower voltage levels.) The corresponding peak effect in $J_c$ is shown in the inset, clearly showing $H_p$ to be 4.30 T. One overall observation from this figure is the fact that all the *V(I)* curves have negative curvature. None have the positive curvature normally associated with thermally activated flow giving way to a liquid-phase flow. In other words, this data does not show any sign of melting from a solid to a liquid phase, a result consistent with the recent SANS study on Nb by Forgan *et al.* [10].

Probably the most striking feature of Fig. 7 is the behavior of the *V(I)* curves at highest currents, something which is easily probed by virtue of the low $I_c$ levels for this sample. There are three cases of note: (1) For $H > H_p$, Fig. 7(a), the curves tend to approach one ohmic straight line, and (2) for $H < H_p$, Fig. 7(b), they also tend to approach a tight set of ohmic straight lines that are widely separated from that of case (1). Finally, (3) For $H = H_p$, the curve approaches a third straight line separate from both cases (1) and (2). These observations are more clearly illustrated in Fig. 8. When compared to constant-current resistance-versus-temperature, $R(T)$, measurements, the level approached by $H > H_p$ curves correspond to that of normal-state resistivity which has a negligible field dependence in this regime. As for $H < H_p$, the fact that the approach is to significantly lower levels signifies that the vortices are not being driven to the normal state but rather to a state of dissipative flux flow. In addition, there is significant field dependence on these asymptotic levels $\rho_f(H, 14.0\ \mathrm{K})$, immediately suggesting BSFF, equation (1).

Using normal state resistivities $\rho_n(T)$ obtained from *RT* measurements, the ratio $\rho_f/\rho_n$ at various constant temperatures are plotted versus $H / H_{c2}$ in Fig. 9. Also shown for comparison are lines corresponding to $C$ values in equation (1) of 1, 1.1, and 1.2, as well as the normal-state-resistivity level. For this range of constant temperatures, the curves actually coincide over a wide range, confirming that the same mechanism is at work in all cases. Furthermore, there is an asymptotic approach to a constant-$C$ BSFF field dependence at lower fields, where presumably the effect of elasticity would be more significant than pinning effects. Meanwhile, at higher fields, $C$ shows some field dependence, causing a downturn of the instantaneous slope, indicating an increasing difficulty in achieving full BSFF and/or the necessity of an even larger driving current level. There is an especially sharp drop just below the peak-effect field $H_p$, again indicating a large departure from ordered flow at $H_p$ although $V(I)$ remains ohmic.

The fact that, at $H_p$, $V(I)$ does indeed become ohmic (Fig. 8) indicates the lowest level of flux flow that can be achieved for this particular temperature. However, distinguishing between scenarios of limited "channel" flow against a background of pinned vortices [22,24] and collective viscous flow is beyond the scope of the present study. What is clear is the following: (1) A stable flux-flow state could not be achieved above $H_p(T)$; (2) a highly ordered vortex state can be achieved only below $H_p(T)$ but with increasing difficulty towards $H_p(T)$; and (3) aside from a gradual current-driven transition from disordered flux flow to highly ordered flux flow, no melting is discernible in this regime.

## IV. CONCLUSIONS

The *H-T* phase diagram in Fig. 10 serves as a summary combining transport data (solid symbols), magnetic data (open symbols), and observations from SANS. The points of the "$H_{irr}$" line was taken from the *M(T)* data, as in Fig. 2, as the point where *M* loses almost all irreversibility at low fields. The $H_{c2}(T)$ line was obtained from both *M(T)* data – in the Fig. 2 lower inset, where magnetization becomes constant – and in-field resistance-versus-temperature *R(T)* data, marking the onset of the superconductive drop in *R*. The peak-effect line $H_p(T)$ was obtained from transport $J_c(H,T)$ – such as the inset of Fig. 7 – and magnetization $J_c(H,T)$, such as the lower plot of Fig. 1. The magnetic and transport data show good consistency. Block arrows signify the presence of long-range order in the regime below $H_p(T)$, competing with the disordering effect of pinning which becomes more relatively significant towards $H_p(T)$. We have shown that this disorder can be minimized to a small extent by field-cooling and to a greater extent by a "shaking" the vortices with a large enough driving pulsed current, wherein we observe Bardeen-Stephen flux flow. As $H_p$ is approached, however, this ordered flux flow becomes more difficult to achieve, and becomes impossible above $H_p$. In the small region between the $H_p$ and $H_{c2}$ lines there is no discernible static or dynamic structure in the vortex matter, signifying a glass or amorphous phase with a field- and current-driven transition to the normal state, Fig. 7(a). All observations are consistent in their confirmation of the Pippard-Larkin-Ovchinnikov picture of the peak effect line $H_p(T)$ as an disorder-driven transition line due to the crossover between the elastic energy of vortices and the pinning energy intrinsic to the superconductor.

***Acknowledgments***

Oak Ridge National Laboratory is managed by UT-Battelle, LLC for the U.S. Department of Energy under contract DE-AC05-00OR22725. The authors also thank H. R. Kerchner, E. Y. Andrei, Z. L. Xiao, E. Zeldov, and Y. Fasano for helpful discussions.

*FIGURE CAPTIONS*

**Figure 1**

Magnetization versus applied field. (a) Hysteresis loop for $T$ = 13.2 K, showing small loop at higher fields, corresponding to the subtle peak effect in $J_c$ in (b) which also shows peaks for two other temperatures. The paramagnetic component in (a) is due to the normal-state Pauli paramagnetism of the V-based system.

**Figure 2**

Magnetic moment versus temperature: upper curve is field-cooled, lower is zero-field-cooled. Lower inset shows the subtle hysteresis close to $H_{c2}$; upper inset shows the corresponding hysteresis proportional to the peak in critical current density $J_c$.

**Figure 3**

Small-angle neutron scattering data as a function of temperature, in constant magnetic field, 6 T, where peak effect temperature from magnetic and transport measurements is 13 K. Arrows indicate history of temperature. (a) Intensity of principal Bragg spot for cooling (circles) followed by warming (squares). Intensity units are of total neutron count. (b) Width of the Bragg spot intensity peak along azimuthal and radial directions, for cooling (solid circles) and warming (open circles). Width measurement is half the width at half of maximum (hwhm) of the peak.

**Figure 4**

Voltage-current ($V(I)$) curves for $T < T_p(H = 4.3\text{ T}) = 14.0$ K, showing a higher-$I_c$ metastability curve (open squares) and the reversible, lower-$I_c$ stable curve (solid squares). See text for details. Dashed straight line indicates the ohmic regime at higher currents. Insets show $V(I)$ for other temperatures at the same field: Top shows $T$ below but closer to $T_p$ where a metastable curve still exists; bottom shows $T > T_p$ where the curves are stable.

**Figure 5**

Driving from metastable to stable *V(I)* curve for $T < T_p$(4.3 T) by stopping at increasingly higher currents $I_s$ (see text). As in Fig. 4, dotted diagonal line marks an ohmic regime at high currents. Dashed horizontal line shows criterion used for $I_c$. Inset: $I_c$ of these *V(I)* curves as a function of $I_s$.

**Figure 6**

Driving from metastable to stable *V(I)* curve for $T < T_p$(4.3 T) by holding at a constant current level of (a) 0.2 A and (b) 0.5 A, for time durations shown. See text for details. For comparison, the same measurement for a stable curve is also shown (upper curves). Inset shows *V* as a function of time as current level is held, for both initially stable and metastable curves.

**Figure 7**

Stable transport *V(I)* curves at constant temperature, for magnetic field *H* (a) $\geq H_p$ and (b) $\leq H_p$. Block arrow shows direction of increasing *H*. Bold *V(I)* curve corresponds to $H = H_p = 4.3$ T. Inset shows corresponding peak effect in $J_c$. The upper dashed curves in (b) are the highest-*H* curves from (a). The lower dashed straight line corresponds to a separate, lower ohmic level, as in Figs. 4 to 6, representing a bundled set of field-dependent levels (see text and Fig. 8).

**Figure 8**

*E/J* versus current *I* plots of data from Fig. 7 for (a) $H \geq H_p$ and (b) for $H \leq H_p$. Arrows indicate direction of increasing *H*, where increments $\Delta H$ are in 0.1 T for $H > 4.30$ T and $H < 4.20$ T. The two sets approach two different high-current resistivity levels *E/J*, with field-dependent levels for (b); note that for $H = H_p$, there is an approach to a third level as well. (See text.)

**Figure 9**

Plot of ratio of flux flow resistivity and normal-state resistivity $\rho_f/\rho_n$ versus $H/H_{c2}$. Included are plots of Equation (1) with prefactor values $C$ = 1, 1.1, and 1.2. Shaded band indicates range of $H = H_p$ for the temperatures shown; dotted horizontal line indicates level for $\rho_n$. (See text.)

**Figure 10**

Magnetic phase diagram, showing data from transport measurements (filled symbols) and magnetization measurements (open symbols). $H_p$ marks the peak effect in $J_c$; "$H_{irr}$" the onset of near-reversibility. See text for details. Block arrows indicate existence of long-range order in the vortex lattice, competing with pinning in the intermediate regime. Direction shows the resulting decrease of order towards higher temperatures for a virgin field-cooled lattice, as well as the increasing difficulty by which disorder can be driven out.

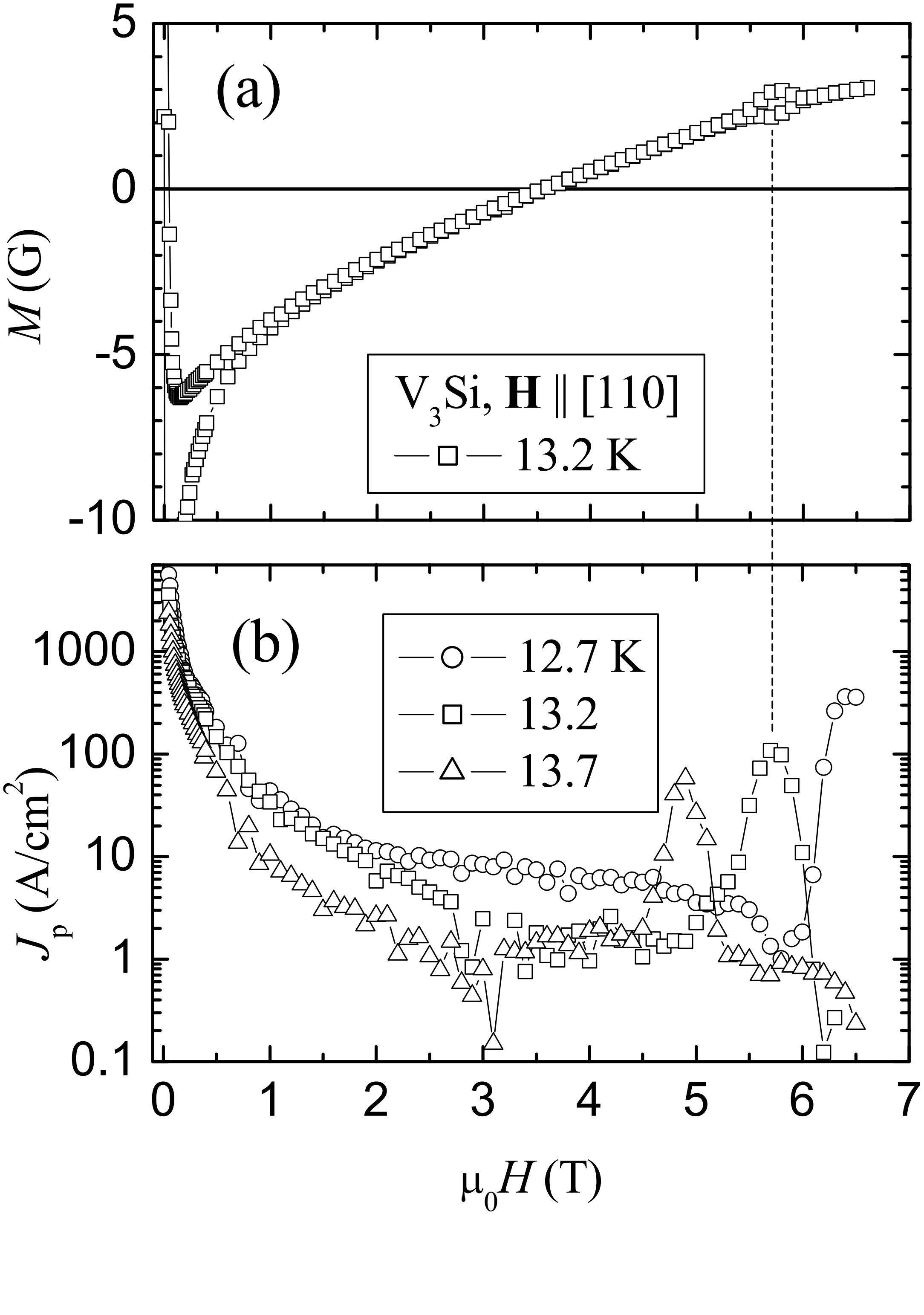

Fig. 1
Gapud et al.

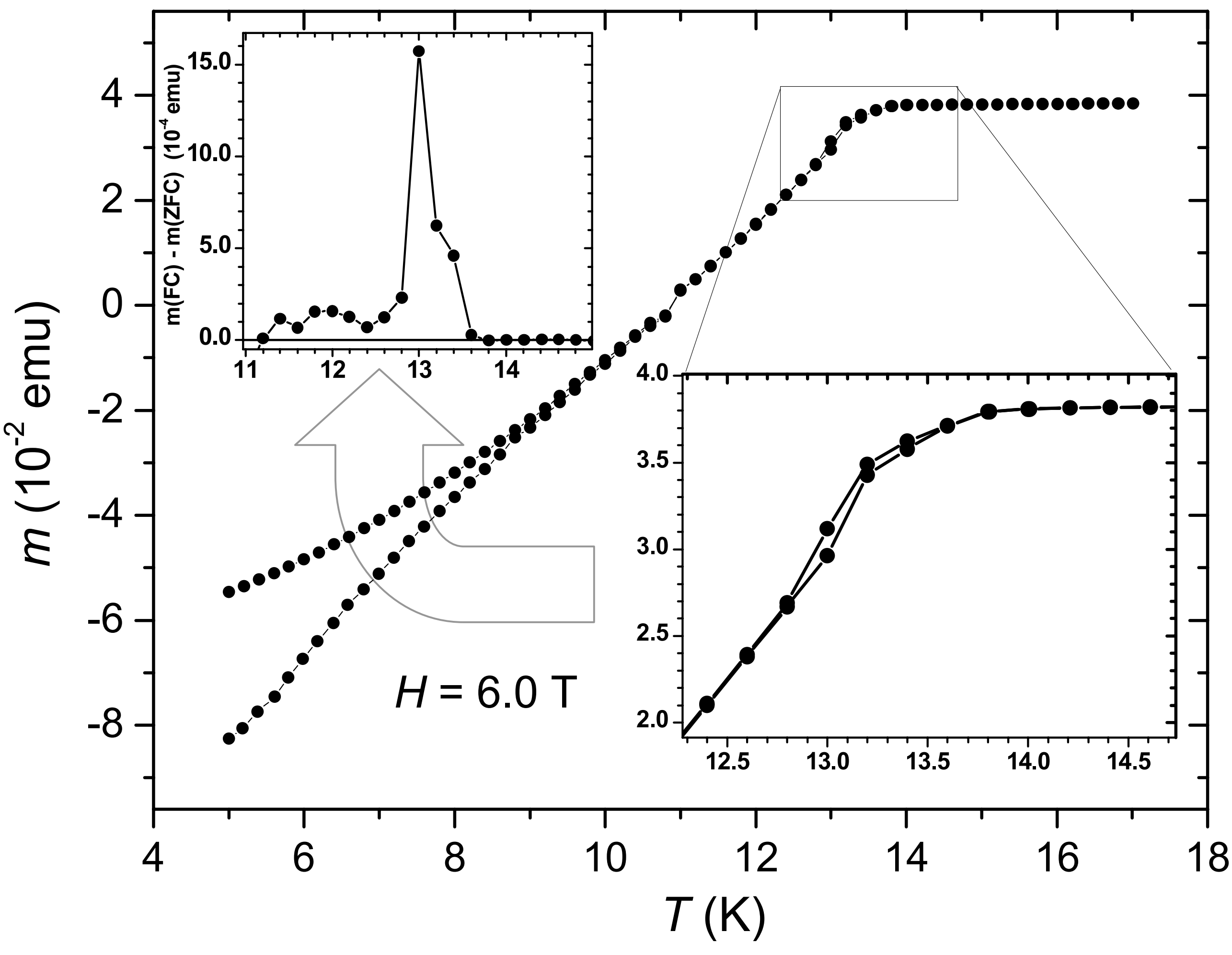

Fig. 2
Gapud et al.

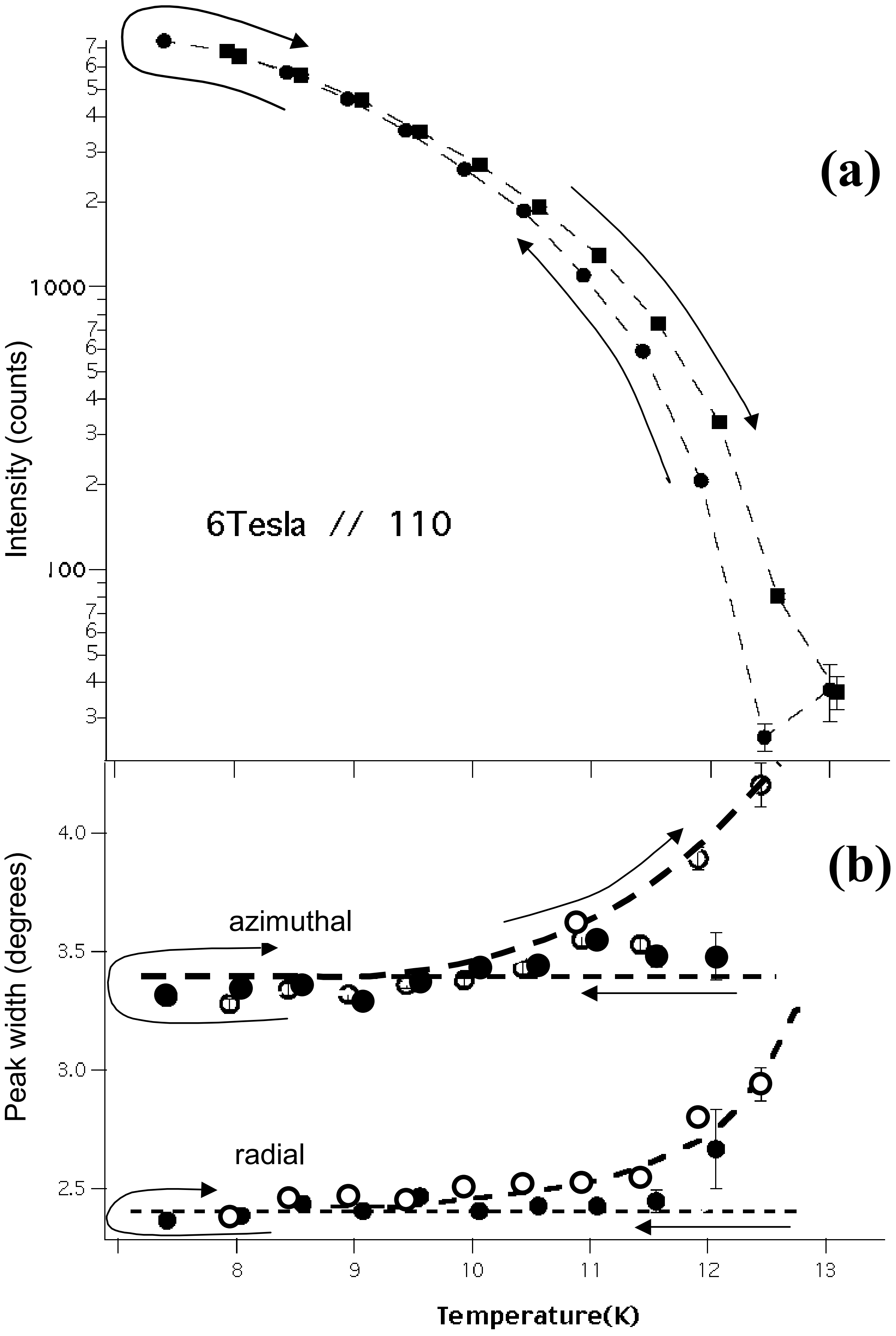

Fig. 3
Gapud et al.

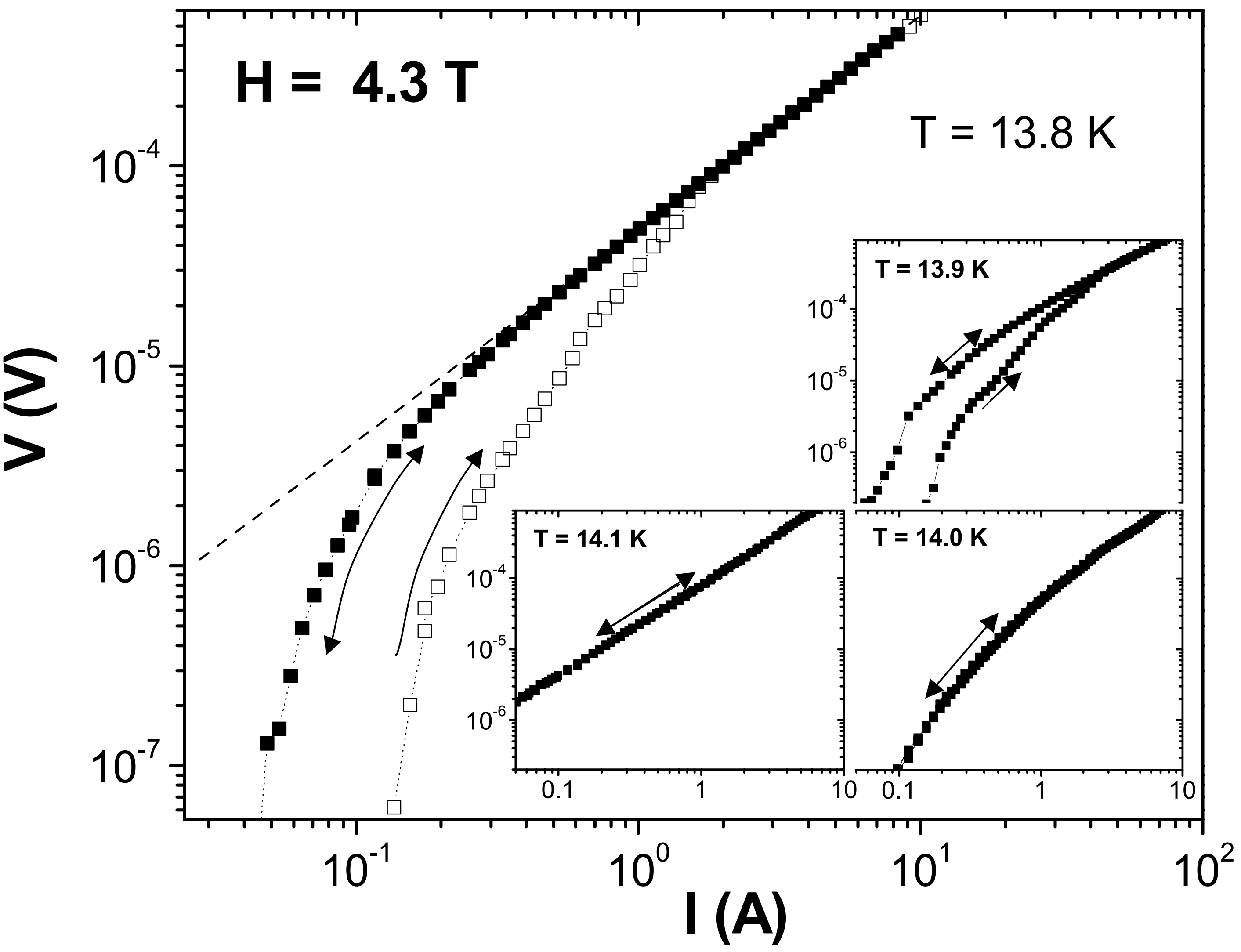

Fig. 4
Gapud et al.

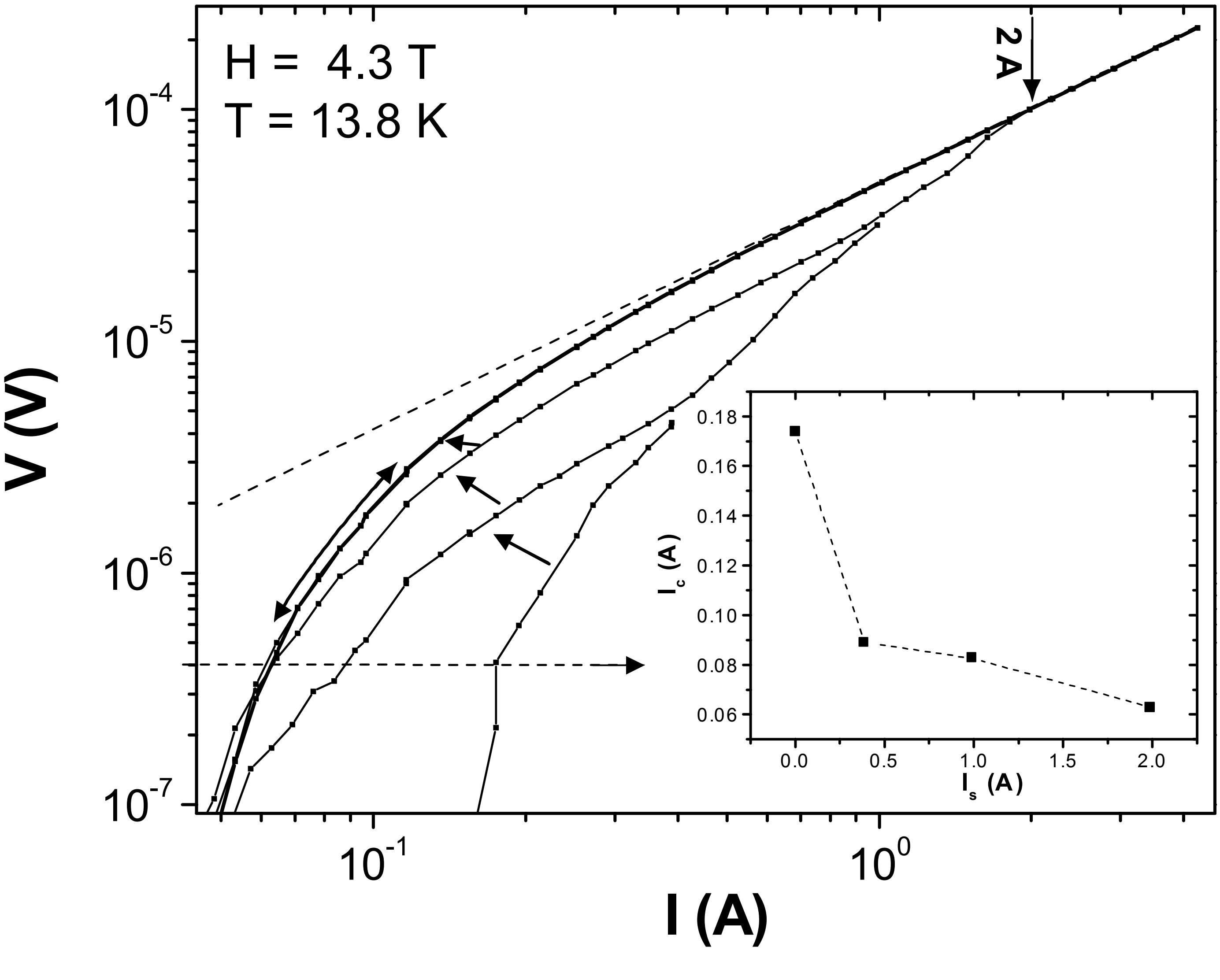

Fig. 5
Gapud et al.

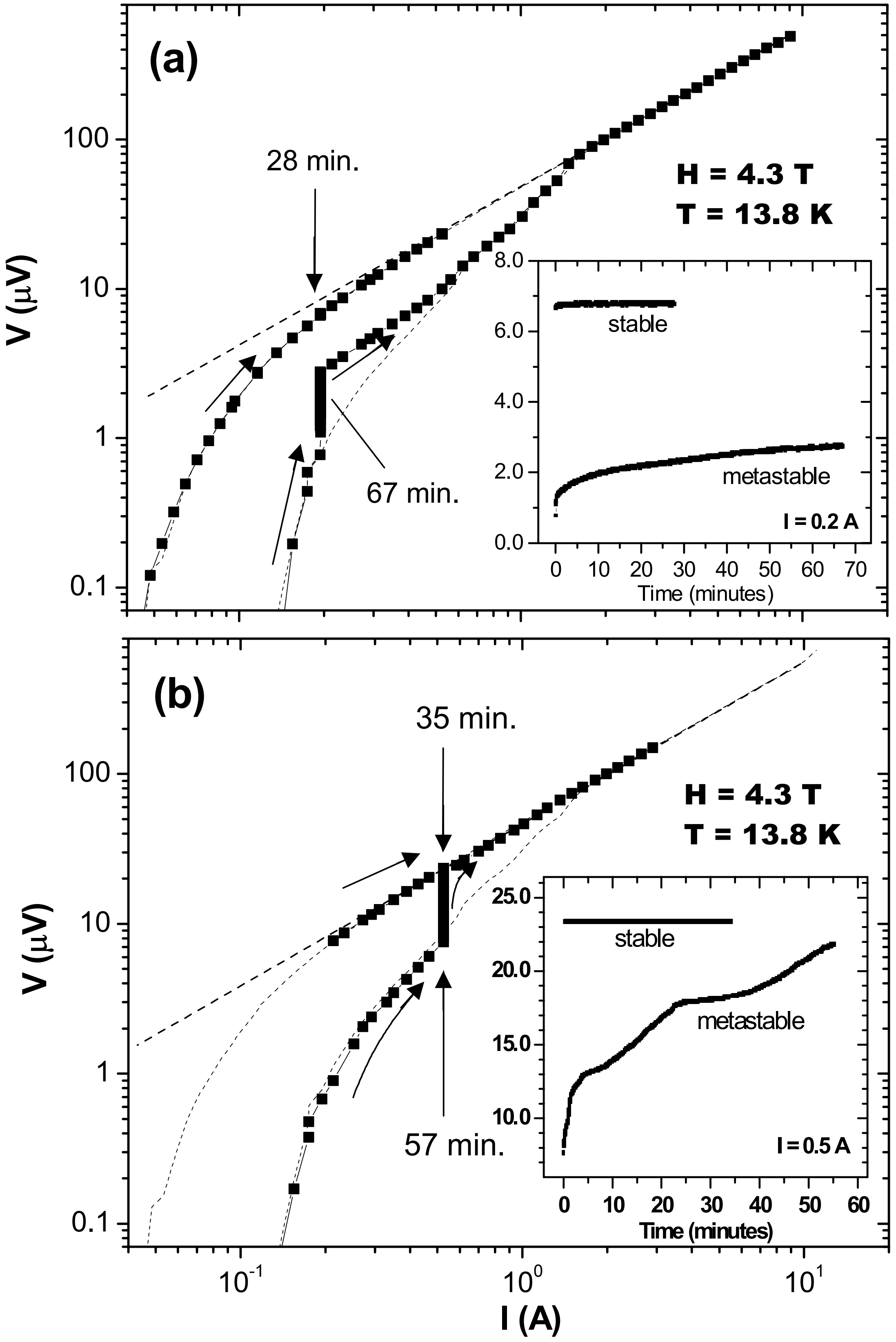

Fig. 6
Gapud et al.

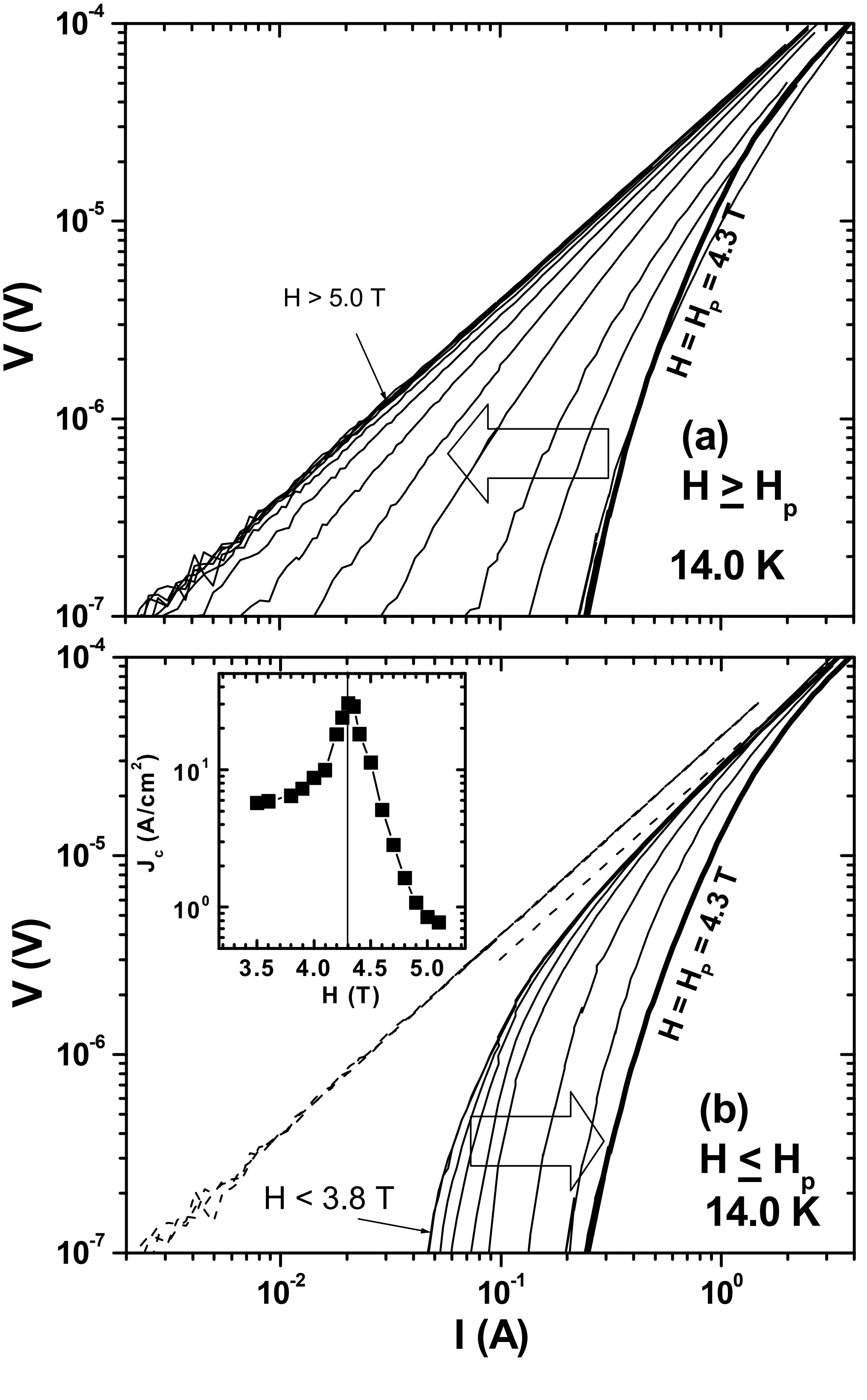

Fig. 7
Gapud et al.

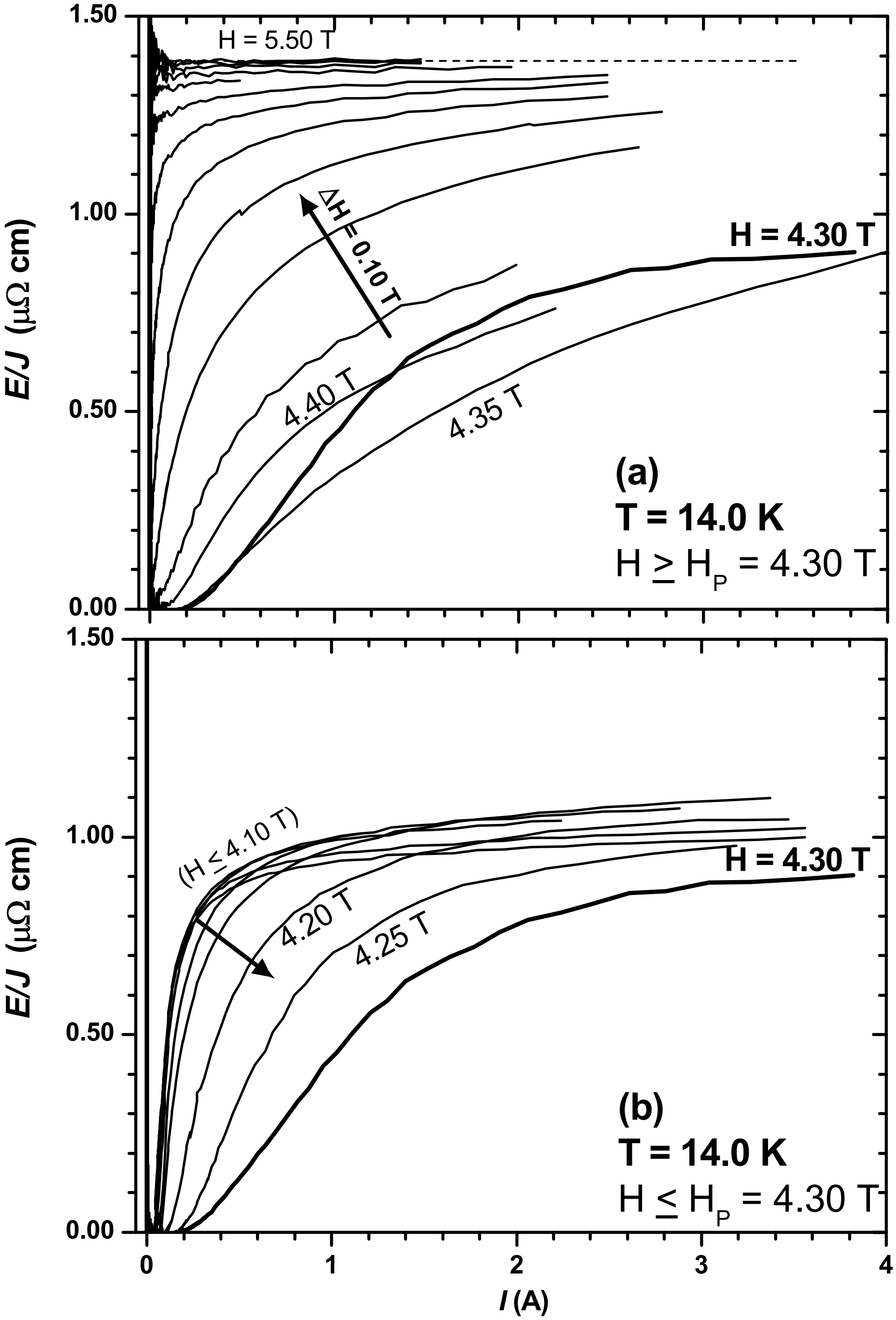

Fig. 8
Gapud et al.

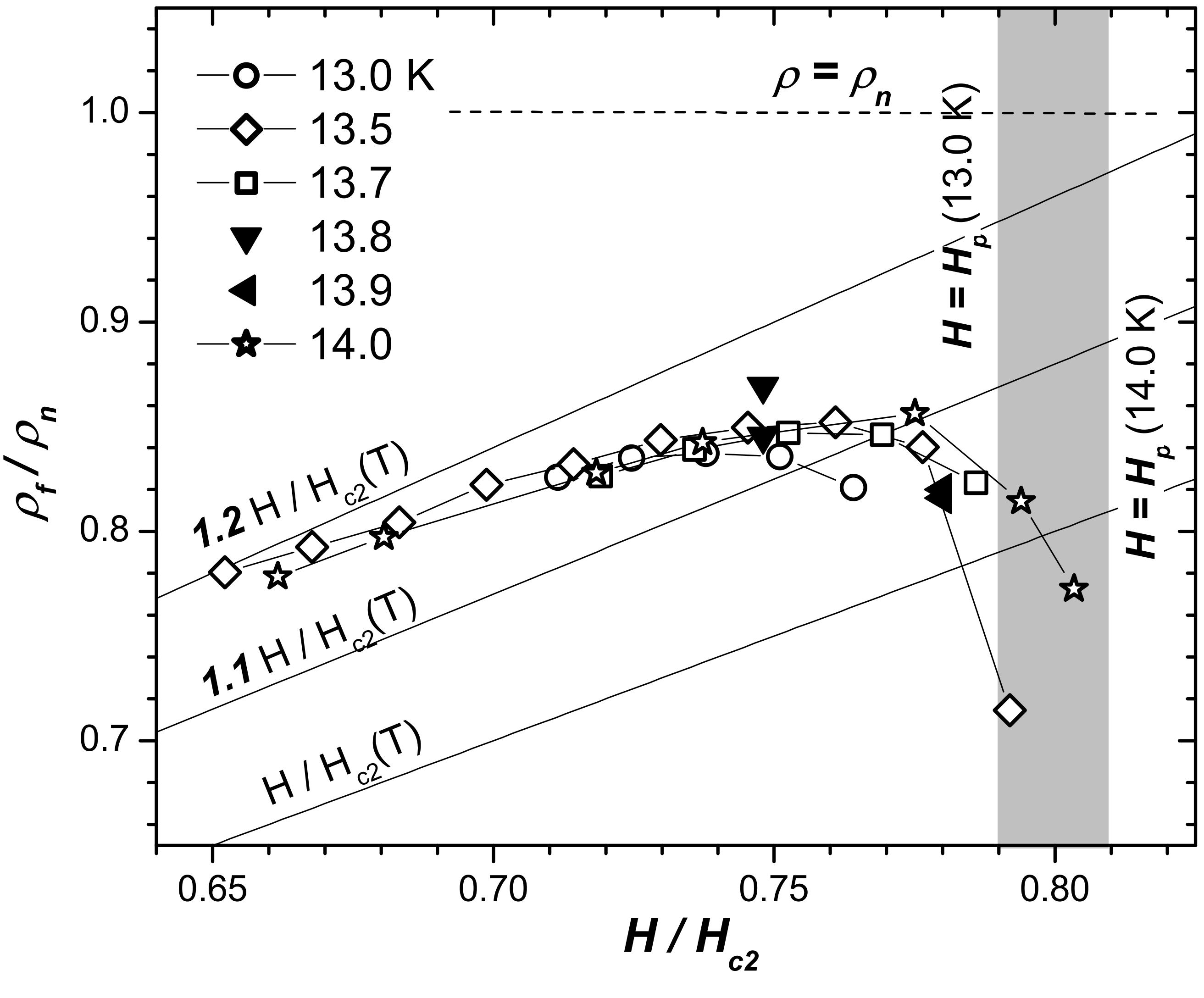

Fig. 9
Gapud et al.

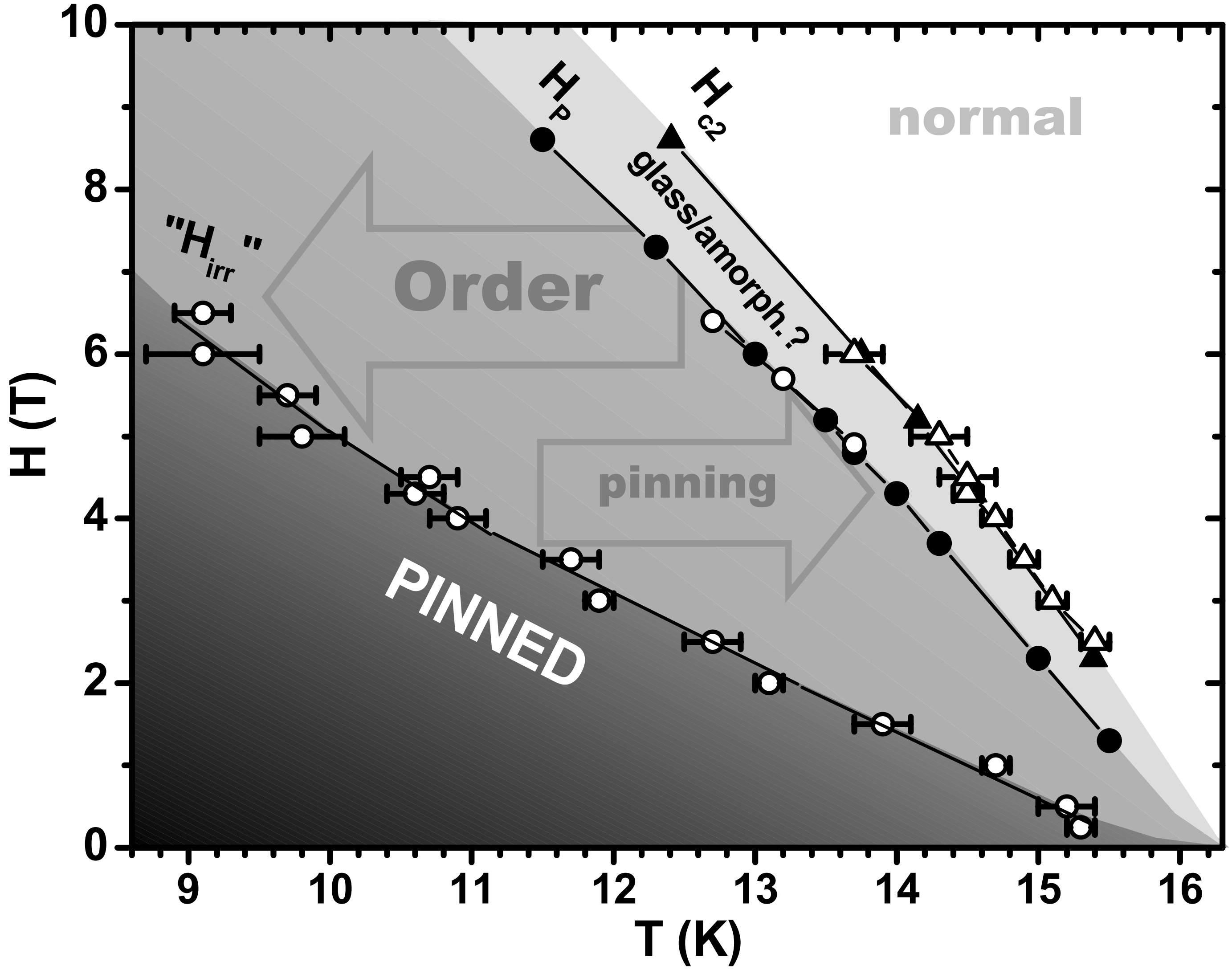

Fig. 10
Gapud et al.